\documentclass[final,5p,times,twocolumn]{elsarticle}

\usepackage{amssymb}
\usepackage{amsmath}
\usepackage{amsthm}
\usepackage{cases}
\usepackage{xcolor}
\usepackage{subfig}
\usepackage[hypcap=false]{caption}
\usepackage{booktabs}
\usepackage{comment}
\usepackage{hyperref}
\usepackage[normalem]{ulem} 

\newcommand{\quotes}[1]{``#1''}
\newcommand{\dint}{\mathop{}\!\mathrm{d}}

\begin{document}

\begin{frontmatter}

\title{Scaling of highly excited Schr\"odinger-Poisson eigenstates \\ and universality of their rotation curves}

\author[1]{Gaia Marangon\corref{cor1}}
\ead{marangon@math.unipd.it}

\author[1,3]{Antonio Ponno}
\ead{ponno@math.unipd.it}

\author[1,3]{Lorenzo Zanelli}
\ead{lzanelli@math.unipd.it}

\cortext[cor1]{Corresponding author.}

\affiliation[1]{
    organization={Department of Mathematics ``Tullio Levi-Civita", University of Padova},
    addressline={Via Trieste, 63}, 
    city={Padova},
    postcode={35131}, 
    country={Italy}}


\affiliation[3]{
    organization={Padua Quantum Technologies Research Center (QTech)},
    addressline={Via Gradenigo, 6/A}, 
    city={Padova},
    postcode={35131},
    country={Italy}}
    
\begin{abstract}
This work provides a comprehensive numerical characterization of the excited spherically symmetric stationary states of the Schr\"odinger-Poisson problem. 
Through numerical computation of highly excited eigenstates, novel heuristic laws are proposed, which describe how their fundamental features scale with the excitation index $n$. 
Key characteristics of the eigenfunctions include: the effective support, which exhibits a parabolic dependence on the excitation index; the distances between adjacent nodes, whose pattern varies regularly with $n$;  and the oscillation amplitude, which follows a power law with an exponent approaching $-1$ for large $n$. 
Based on the eigenfunctions, eigenvelocities are conveniently defined. They exhibit a mid-range oscillatory region with an average linear trend, whose slope approaches zero in the large $n$ limit; and they are characterized by heuristic scaling relationships with the excitation index $n$, revealing an intrinsic universal behavior. 
\end{abstract}



\end{frontmatter}


\section{Introduction} \label{sec:introduction}
The Schr\"odinger-Poisson model, or Schr\"odinger-Newton model, embeds a nonlinear, nonlocal effect in a Schr\"odinger dynamics through the coupling with a Newtonian potential. It describes the evolution of a scalar matter field $\psi(t,x)$ via the  Schr\"odinger equation, where $|\psi|^2(t,x)$ represents the matter density distribution of the field. The matter field is subjected to self-gravity, which is modeled through a Poisson potential $\phi(t,x)$, with the density $|\psi|^2(t,x)$ acting as the source. In dimensionless form, the system reads:
\begin{align}
\begin{cases}
    i\partial_t \psi(t,x) = (-\Delta +2\phi(t,x)) \psi(t,x)\\
    \Delta \phi(t,x) = |\psi|^2(t,x) 
\end{cases}   
\qquad \text{with } x \in \mathbb{R}^3 \,.
\label{eq:SP}
\end{align}

This system is relevant in a wide variety of fields \citep{Paredes2020}. It was first introduced by \citet{Ruffini1969} as the non-relativistic approximation of general relativistic, self-gravitating systems of bosons. It was subsequently applied in astrophysics and cosmology to study non-relativistic boson stars \citep{Schunck2003} or to model ultralight scalar field dark matter (\citep{Matos2024,Hui2017}). Interpreting the quantum nature of the system at a more fundamental level, \citet{Diosi1984} and \citet{Penrose1998} used the Schr\"odinger-Poisson model to explore the effects of self-gravitation on a single particle wavefunction, while a more recent research line \citep{Penrose2014,Bahrami2014} investigates the problem of quantum gravity, showing that a Schr\"odinger-Poisson nonlinearity emerges naturally from a fundamentally semi-classical interpretation of gravity. Interestingly enough, the implications of this approach might also be tested through optomechanics \citep{Grossardt2016,Bekenstein2015} or through dilute cold atom BECs experiments \citep{Lahaye2009}, where isotropic condensed configurations can be obtained through an electromagnetically induced gravity-like interaction \citep{Giovanazzi2001,Mendonca2019,GarciaRipoll2003}. Other fields of application for the Schr\"odinger-Poisson model include non-linear optics, where it can describe the propagation of light in thermo-optical media \citep{Navarrete2017,Bekenstein2015},  plasmas and electrons in semiconductors or metallic structures (see \citet{Paredes2020} and references therein).

\begin{figure}[htb]
    \centering
    \includegraphics[scale=0.55]{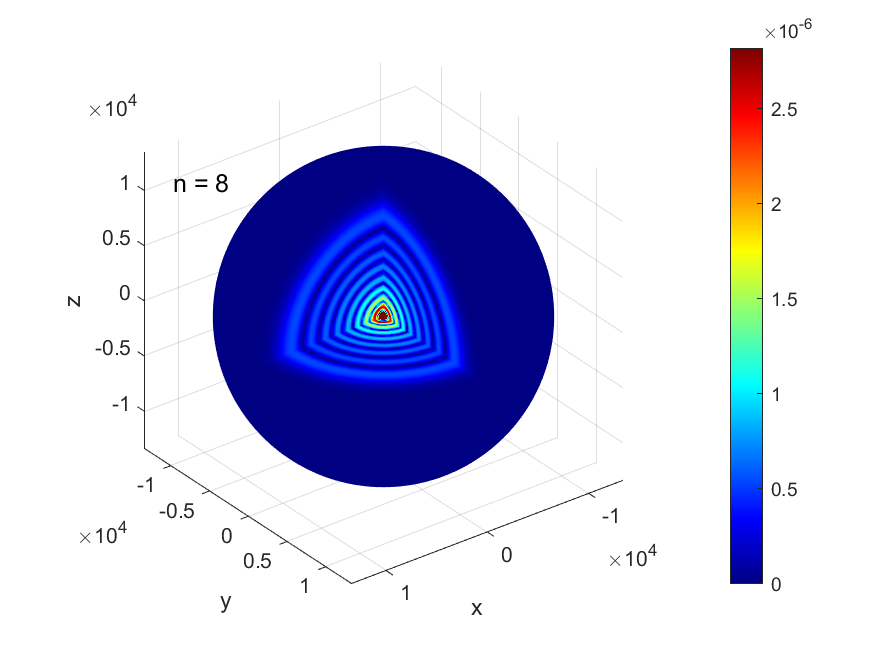}
    \caption{Three-dimensional section of the matter distribution, as predicted by the $8$-th stationary state $f_8(r)$. For visualization purposes, the modulus $|f_{8}|(r)$ is shown rather than the actual predicted matter density $|f_{8}|^2(r)$. The structure exhibits $9$ shells of decreasing density separated by $8$ voids (shown in dark blue).}
    \label{fig:section}
\end{figure}

Despite the wide popularity of the Schr\"odinger-Poisson system, the understanding of its mathematical structure remains incomplete. In this work we focus on radially symmetric stationary states, which are solutions to system \eqref{eq:SP} of the form $(e^{i\varepsilon t}f(r),\phi(r))$. From \citet{Lieb1977} and \citet{Lions1980}, they are known to comprise an infinite, discrete family
$\{\varepsilon_n,\,f_n(r),\,\phi_n(r)\}_{n=0}^\infty$, with positive eigenvalues $\varepsilon_n >0$, real eigenfunctions $f_n(r)$ and excitation index $n \in \mathbb{N}$. 
Currently, no analytic expressions exist for these eigenstates, which are therefore mainly known through numerical computations \citep{Sin1994,Bernstein1998} and asymptotic estimates \citep{Roque2023}.
As shown in Figure \ref{fig:section}, the $n$-th excited stationary state represents matter distributions as $n+1$ concentric shells, separated by voids and exhibiting decreasing density outwards. A rigorous characterization of this structure and its dependence on the excitation index $n$ remains incomplete, which this paper aims to address.

Given a predicted matter density distribution $f_n^2(r)$ , one can calculate the tangential velocity of a test object in circular orbit at radius $r$ from the distribution's center. 
This calculation assumes the velocity is determined by the gravitational force exerted by matter within a sphere of radius $r$:
\begin{align}
    v_n(r) = \sqrt{ \frac{ \int_0^{r} f_n^2(s) s^2 ds}{r}} \,.
    \label{eq:velocity}
\end{align}
This type of prediction, where a velocity profile is derived from the density distribution of a proposed model, is typical of galactic dark matter research, as in this context it can be compared with experimental rotation curves of spiral galaxies \citep{Sin1994,Lelli2016,Katz2016,Harko2022,Guzman2015,Binney2011}, providing both model validation and deeper physical insights into the problem. 
The use of excited Schr\"odinger-Poisson stationary states as models for ultralight galactic dark matter distributions and the approximate behavior of the corresponding rotation curves were  explored in a series of works \citep{Sin1994,Ji1994,Lee1996}, matching some relevant physical expectations \citep{Burstein1985,Rubin1980,Rubin1985}. However, their role was reconsidered after the discovery of numerical instability \citep{Guzman2004,Guzman2006} and the development of more detailed three-dimensional simulations of system \eqref{eq:SP} \citep{Schive2014}. The latter showed the formation of a flat, oscillating core surrounded by an envelope of highly energetic granular structures, which continuously fluctuate in density and velocity. 
The solid angle averages and the time averages of the resulting anisotropic, time-dependent density configurations provide a universal density profile, whose core is well fitted by the ground state of the Schr\"odinger-Poisson problem \eqref{eq:SP} and whose halo resembles the Navarro-Frenk-White profile of standard CDM \citep{Schive2014,AlvarezRios2024}.  
The subsequent literature aimed to construct on-demand density profiles describing such average behaviors and to study their stability with simulations \citep{AlvarezRios2024}, eventually fitting them to experimental data \citep{Lelli2016,Katz2016}.
Such effective density profiles usually involve the ground state of the Schr\"odinger-Poisson problem \eqref{eq:SP}, which behaves as a stable attractor \citep{Guzman2006} (see also \citep{AlvarezRios2024} for a three-dimensional generalization), while a wider variety of choices are adopted to model the average behavior of the granular envelope.
The excited Schr\"odinger-Poisson stationary states are still used in some contemporary descriptions of dark matter density profiles, which assume simultaneous occupation of several states resulting in stable effective multimodal configurations \citep{UrenaLopez2010,Guzman2020,AlvarezRios2024}. 
Despite the nonlinearity of the problem, which complicates the use of knowledge on excited stationary states to study superposed configurations, we believe that a refined understanding of the properties of such states may be of interest, independently of the specificities of the possible applications.
Let us also mention that, even though the excited Schr\"odinger-Poisson eigenstates are unstable, some recent results showed that the highly excited ones have a lifetime that increases with $n$
\cite{Roque2023} which, in view of the broad range of physical applications in which the Schr\"odinger-Poisson model is applied, opens the possibility for further analysis on this topic. A comprehensive study on the stability issue, whose interest goes beyond the specific application to the dark matter problem, is out of the scope of this work and we just include some remarks in the conclusive Section of this paper, postponing a more thorough discussion to future works. 

Building on these considerations, we include in our study a comprehensive, quantitative description of the eigenvelocities \eqref{eq:velocity} arising from the Schr\"odinger-Poisson model, characterizing their key features as explicit functions of the excitation index $n$.

\subsection{Comparison with previous works and main results}
From an analytical perspective, pioneering work on Schr\"odinger-Poisson stationary states was conducted by \citet{Lieb1977}, who established the well-posedness of the ground state, and \citet{Lions1980}, who first described excited stationary states.
Numerical investigations of low-excitation states ($n \lesssim 15$) were later performed by \citet{Bernstein1998,Moroz1998,Harrison2003}, providing significant qualitative insights into eigenfunction properties. These eigenfunctions display oscillatory behavior, featuring $n$ nodes and $n+1$ local extrema of alternating sign with diminishing amplitudes, decaying monotonically to zero beyond the last oscillation.
To address the lack of exact analytical expressions, alternative density functions such as $\rho(r) = \sin(r)r^{-1}$ were proposed (see \cite{Matos2024,Bohmer2007}). While these approximations are widely adopted in physical applications, they are inadequate to capture both the irregular nodal spacing and the precise power law decay of the eigenfunctions, highlighting the need for deeper structural understanding.

Subsequent work provided partial analytical support for these numerical observations.  \citet{Tod1999} demostrated the smoothness and boundedness of the eigenfunctions, while \citet{Tod2001}, focused on bounds, proving that all eigenvalues are negative. 
\citet{Greiner2006} established connections  with quantum defect theory, and \citet{Kiessling2021} investigated the long-range behavior of the eigenfunctions.

However, a comprehensive analysis of how properties scale with the excitation index $n$ remains lacking. To address this gap, we propose quantitative, heuristic laws derived from numerically computed highly excited stationary states ($n \le 80$), aiming to provide foundations for future analytical and physical studies.
Specifically, we analyze the eigenfunctions' structure, characterizing them as oscillatory functions modulated by a power law. 
The radial position of the last oscillation, which approximately defines the eigenfunction support, exhibits a clear parabolic dependence on $n$, characteristic of Keplerian problems like the Bohr model for the hydrogen atom (see e.g. \cite{Griffiths2017}).
The spacing between adjacent nodes increases with the radial position, following a consistent pattern across increasing $n$ values -- showing modest increases at small radii and more pronounced increases at larger radii. This behavior constitutes a key correction to the $\sin{(r)} r^{-1}$ approximation, which fails to account for irregular nodal spacing.
Finally, the power law governing amplitude modulation is characterized by fitting local extrema at small radial positions, with deviations observed near the last extremum. The power law exponent approaches $-1$ asymptotically as $n$ increases, recovering the $r^{-1}$ approximation in the large $n$ limit. 

From a physical perspective, rotation curves are used in the context of galactic dark matter distributions as the primary experimental reference to compare with analytical predictions. In this framework, observed rotation curves typically show mid-range oscillatory behavior with an approximately linear trend, forming a plateau. This plateau generally exhibits a near-flat slope, although both positive and negative slopes have also been documented (see e.g. \citep{Rubin1980,Sofue2015,Lelli2016}). Moreover, physical intuition suggests that galactic-scale dark matter clusters, which generate these rotation curves, might exhibit some kind of universal behavior \citep{Rubin1985,Burstein1985}, possibly in an average sense \citep{AlvarezRios2024}.
\citet{Sin1994} first noted that the key features displayed by the eigenvelocities in the Schr\"odinger-Poisson model align well with these physical observations, especially when taking into account the presence of a baryonic bulge \citep{Ji1994}.
As already discussed, the interpretation of these similarities is to be intended with the due caution. In view of the possible role of the excited stationary states in contemporary models, as the multimodal ones, we still add some comments, refining the observations on the mathematical structure of the eigenvelocities and emphasizing in particular the large-$n$ asymptotic behaviors, for which the instability issue might have a reduced impact \citep{Roque2023}.
We note that the large-$n$ approximation $f_n(r) \sim r^{-1}$ derived from the local extrema fit implies an approximately constant eigenvelocity \eqref{eq:velocity}, consistent with an expected flat plateau. Our systematic results refine this intuition, showing how the slopes of the mid-range eigenvelocities decay with $n$ according to a power law.
In addition, the velocity at the last local oscillation exhibits a clear dependence on its radial position, which itself varies with $n$. This relationship reveals appropriate scaling laws under which all predicted eigenvelocities converge to a similar shape, independent of $n$, exhibiting an intrinsic universal behavior that might be useful in modeling multimodal configurations with average common features.

\section{Model} \label{sec:model}
The Schr\"odinger-Poisson model \eqref{eq:SP} describes the dynamics of a density $|\psi|^2(t,x)$ under self-gravity.
The model couples a Schr\"odinger equation for the matter field $\psi(t,x)$ with a Poisson equation for the potential $\phi(t,x)$.
Its stationary states are expressed in the form $(e^{i\varepsilon t}f(x),\phi(x))$, where the phase oscillation in time naturally emerges from the norm conservation law $\partial_t (\int_{\mathbb{R}^3} |\psi(t,x)|^2 \dint^3x) = 0$, with $\varepsilon$ representing the associated energy parameter.  The spatial component $f(x)$ of the matter field can be assumed real without loss of generality.

As noted by \citet{Lieb1977}, the problem exhibits invariance under norm scaling:
\begin{align}
    \tilde{x} &= N^{-1}x \,;&
    \tilde{t} &= N^{-1}t \,;&
    \tilde{f} &= N^2 f \,;&
    \tilde{\phi} &= N^2 \phi \,. 
\end{align}
This introduces a degree of freedom in defining stationary states, which is resolved by selecting a specific value for the $L^2(\mathbb{R}^3)$-norm of $f$. Including this normalization constraint and explicitly solving the Poisson equation, the stationary problem becomes:
\begin{subnumcases}{\label{eq:ChoquardPoisson}}
    \Delta f(x) +\frac{1}{2\pi}\left(\int_{\mathbb{R}^3} \frac{f^2(y)}{|x-y|} \dint^3y \right) f(x) = \varepsilon f(x)
    &\label{eq:Choquard}\\
    \phi(x) = - \frac{1}{4\pi}\int_{\mathbb{R}^3} \frac{f^2(y)}{|x-y|} \dint^3y 
    &\label{eq:solvedPoisson} \\
    \int f^2(x) \dint{^3 x}= 1 
    &\label{eq:normalization}
\end{subnumcases}
where equation \eqref{eq:Choquard} is known as the Choquard equation and represents a nonlinear eigenvalue problem, with eigenvalue $\varepsilon$ and eigenfunction $f(x)$.
Following the prevailing approach in the literature, we focus on spherically symmetric solutions, which form an infinite, discrete family $\{\varepsilon_n,\,f_n(r),\,\phi_n(r)\}_{n=0}^\infty$, with excitation index $n \in \mathbb{N}$ \citep{Lions1980}. 
For each eigenfunction $f_n(r)$, we associate the velocity field $v_n(r)$ defined in \eqref{eq:velocity}, which we term eigenvelocity (in the definition, we omit the factor $4\pi$ from angular integration for simplicity).

\subsection{Implementation}
Eigenstates $\{\varepsilon_n,\,f_n(r),\,\phi_n(r),v_n(r)\}_{n=0}^\infty$ are numerically computed up to high excitation indices ($n \le 80$) following the implementation described by \citet{Bernstein1998} (see also \citep{Harrison2003}). The numerical scheme employs two nested iterative procedures: an outer iteration handles the representation of the infinite domain, while an inner iteration solves problem \eqref{eq:ChoquardPoisson} on a fixed domain.
Specifically, the outer iterative procedure progressively extends the computational domain until the eigenvalue correction falls below a specific tolerance. The inner procedure decouples problem \eqref{eq:ChoquardPoisson} by using previous-step approximations (or an initial guess for the first step) for the unknowns $f(r)$ and $\phi(r)$, solving the resulting approximate problem to achieve increasingly accurate solutions.

To verify algorithmic robustness, numerical computations were performed with various spatial grid refinements. This validation process allowed to monitor the quality of the results, confirming reliability up to $n=80$, while results for higher excitation indices were deemed insufficiently accurate for this study and excluded from the analysis.

\section{Heuristic Laws} \label{sec:heuristic}
The numerical implementation of excited stationary states enables a comprehensive quantitative analysis of their structure. 
We first present their qualitative characteristics, identifying key features and establishing notation. Then, we conduct a systematic study, deriving heuristic laws that describe how fundamental properties scale with the excitation index $n$.

\subsection{Eigenfunctions: main features and notation}
\begin{figure}[htb]
\centering
     \subfloat[][{Structure of the eigenfunction showing nodes and extrema.}]{
    \includegraphics[scale=0.35]{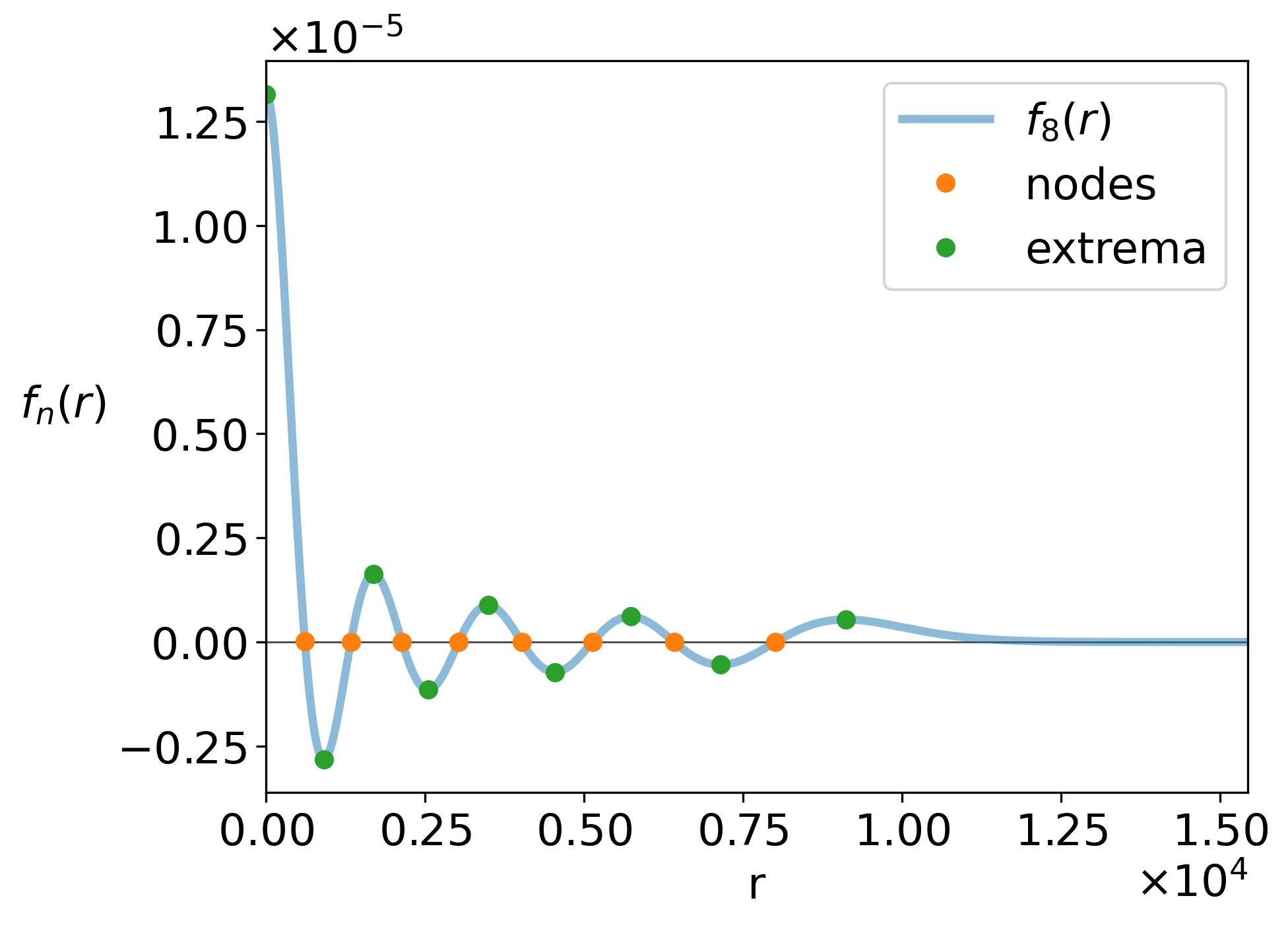}
    \label{fig:fDetails}
    }\\
    \subfloat[][{Detailed view with notation.}]{
    \includegraphics[scale=0.35]{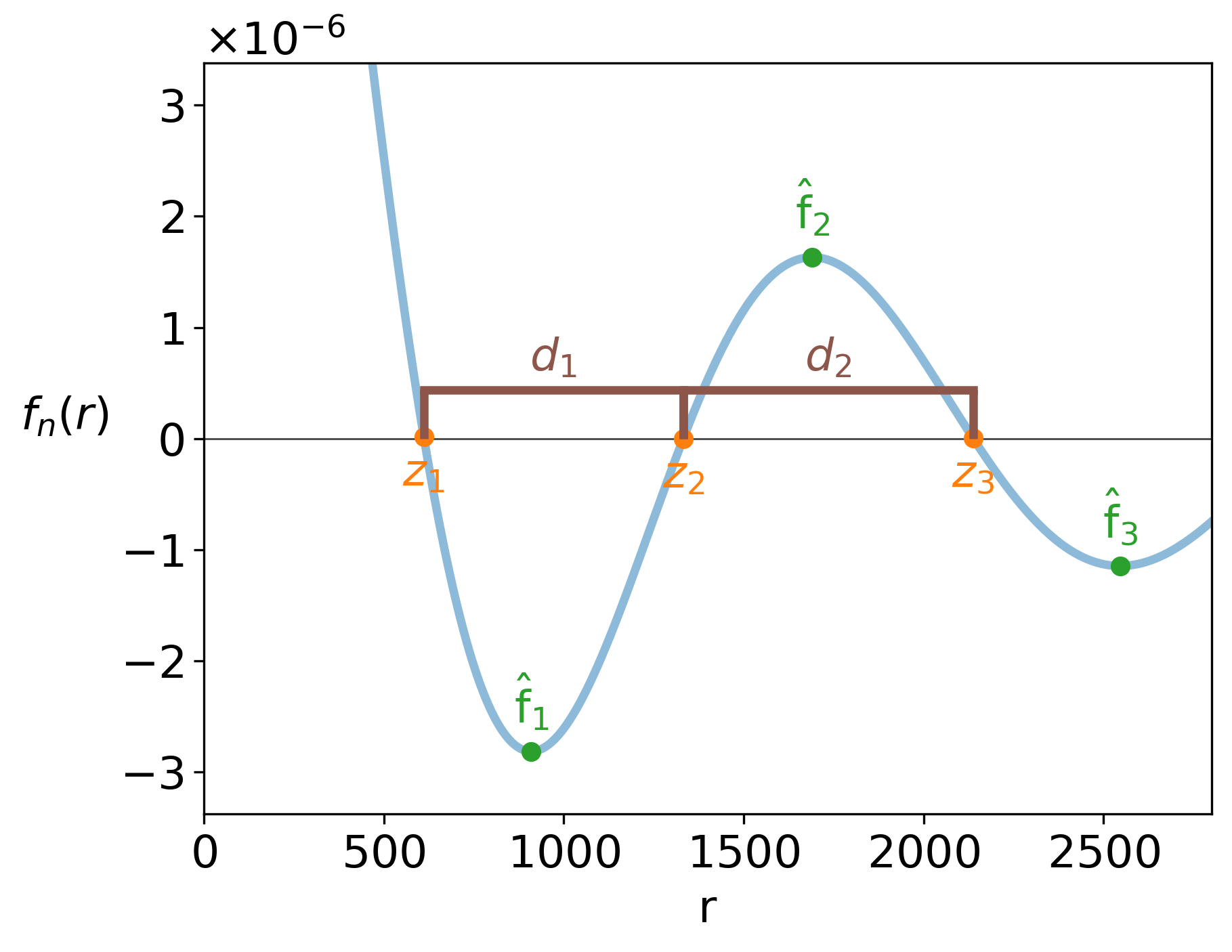}
    \label{fig:fDetails_zoom}
    } 
\caption{Structure and notation of eigenfunction $f_{8}(r)$. Panel (a) shows the complete eigenfunction, with nodes $\{z_i\}_{i=1}^8$ marked in orange and local extrema $\{(\hat{r}_i,\hat{f}_i)\}_{i=0}^8$ in green. Panel (b) provides a magnified view highlighting the notation, including the first two nodal distances $\{d_i\}_{i=1}^{7}$ with $d_i \equiv z_{i+1}-z_{i}$.}
\label{fig:eigenfunction_notation}
\end{figure}

The structure of the $n$-th eigenfunction $f_n(r)$ is illustrated in Figure \ref{fig:eigenfunction_notation}, which states the notation for its key features. For clarity, the figure shows an eigenfunction with a low excitation index ($n=8$), though the structure remains analogous for highly excited eigenstates. 

The function exhibits an oscillatory behavior with $n$ nodes and $n+1$ local extrema. The nodes are defined as the $n$ radial positions $\{z_i\}_{i=1}^n$ where the eigenfunction vanishes, $f_n(z_i)=0$. The $n-1$ intervals between adjacent nodes, denoted as $\{d_i\}_{i=1}^{n-1}$ and defined by $d_i \equiv z_{i+1}-z_{i}$, are termed nodal distances. For each distance $d_i$, we refer to $z_{i+1}$ and $z_{i}$ as the right and left nodes, respectively.

The $n+1$ local extrema, denoted by $\{(\hat{r}_i,\hat{f}_i)\}_{i=0}^n$, alternate in sign and decrease in amplitude. The central point $(\hat{r}_0,\hat{f}_0)$ represents the global maximum of the eigenfunction, while the outermost local extremum $(\hat{r}_n,\hat{f}_n)$ marks the end of the oscillating region, beyond which the eigenfunction decreases monotonically to zero. In physical applications, the radius $\hat{r}_n$ serves as an indicator of the function's effective support. 

When applying these eigenstates to model matter fields, the squared modulus $|f_n|^2(r)$ represents the matter density distribution. Its three-dimensional structure, revealed by numerical computations and shown in Figure \ref{fig:section}, describes matter as arranged in concentric spherical shells of decreasing density.
The quantitative study that follows characterizes this structure by deriving precise heuristic laws, that describe its dependence on the excitation index $n$.

\subsection{Approximate support}
\begin{figure}[htb]
    \centering
    \includegraphics[scale=0.35]{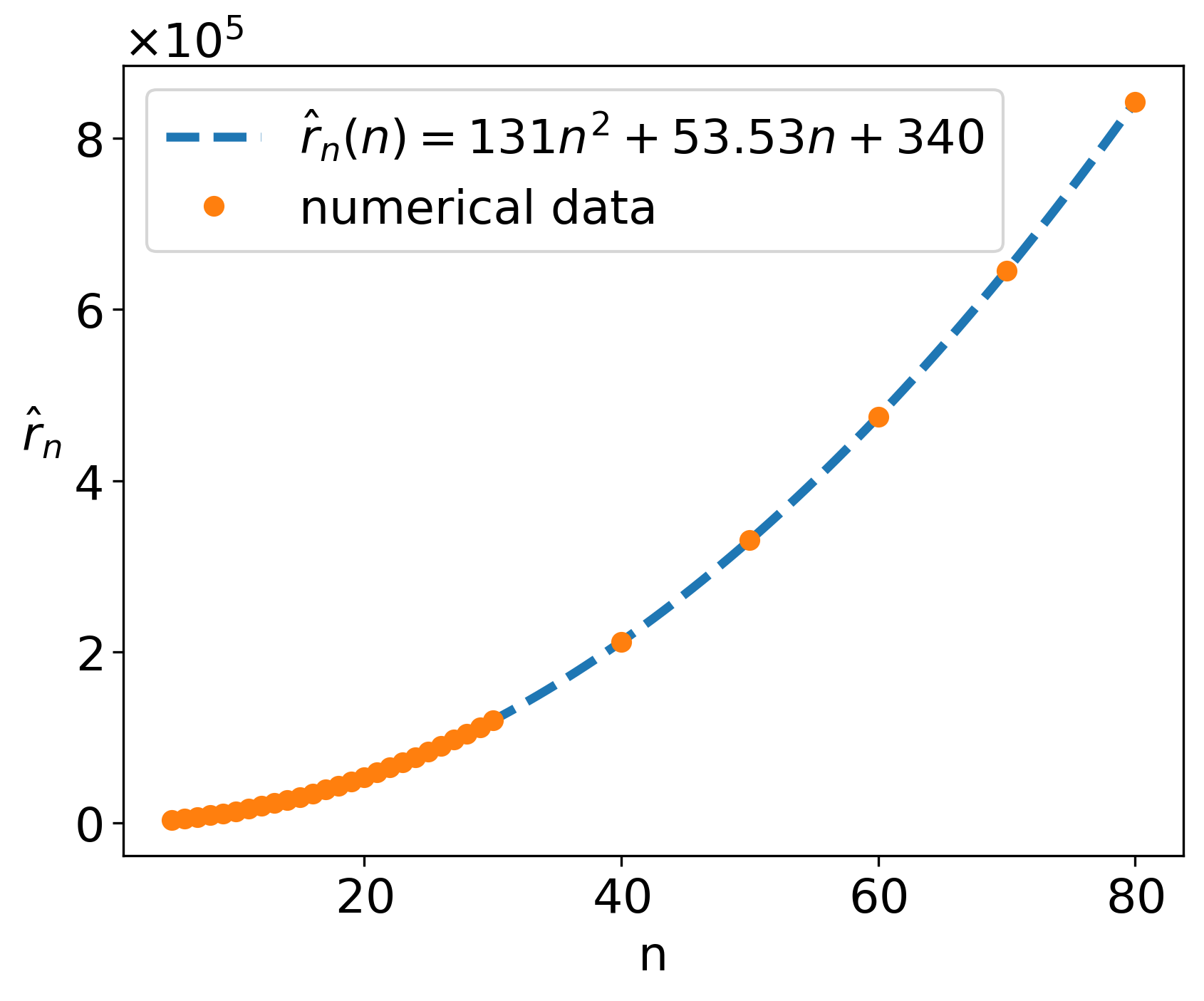}
    \caption{Parabolic fit for the radial position $\hat{r}_{n}(n)$ of the outermost local extremum, serving as an indicator of the function's support.}
    \label{fig:rout_fit}
\end{figure}

Let us examine the radial position $\hat{r}_{n}(n)$ of the outermost local extremum, which serves as an indicator of the function's support in physical applications. We computed its value for several excitation indices $n$, with the results displayed in Figure \ref{fig:rout_fit}. The data reveals a clear parabolic relationship, resulting in the following numerical fit: 
\begin{align}
    \hat{r}_{n}(n)=131n^2 + 53.53n + 340 \,.
    \label{eq:rout_fit}
\end{align}
This heuristic law provides valuable physical insight by describing how the effective support of the matter distribution $f_n^2(r)$ varies with the excitation index $n$. Note that the asymptotic scaling $\hat{r}_{n}(n) \sim n^2$, observed at large $n$ is characteristic of the quantum Keplerian or hydrogenoid problems (see \citet{Griffiths2017}).  

\subsection{Nodal distances}
\begin{figure}[htb]
    \centering
    \includegraphics[scale=0.35]{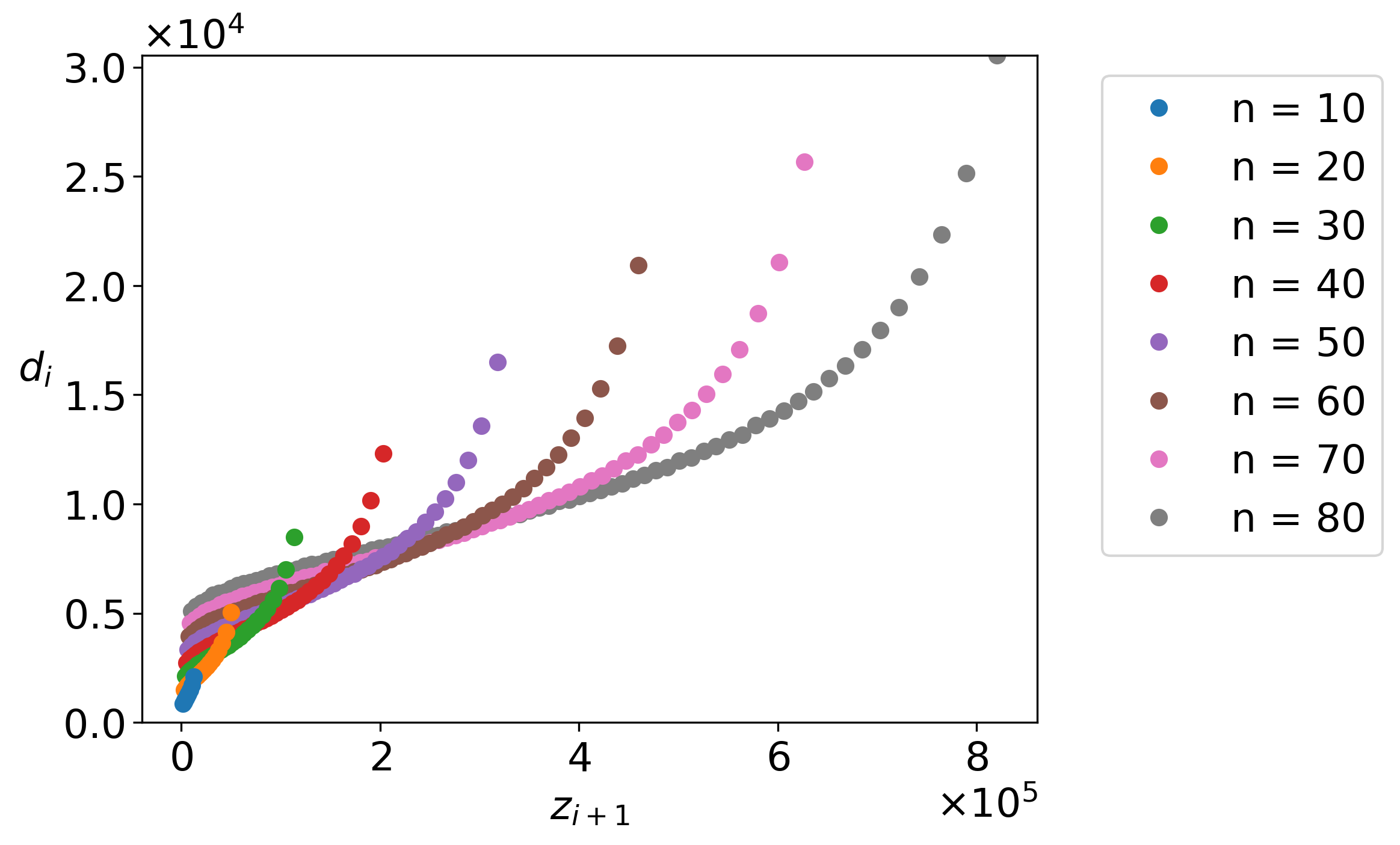}
    \caption{Nodal distances $\{d_i\}_{i=1}^{n-1}$ plotted against their corresponding right nodes $\{z_{i+1}\}_{i=1}^{n-1}$, for several sample eigenfunctions.}
    \label{fig:domain_nodalDistance}
\end{figure}

Let us then examine the nodal distances $\{d_i\}_{i=1}^{n-1}$ between adjacent nodes. As Figure \ref{fig:eigenfunction_notation} suggests, these distances appear to increase with radial position. To investigate this observation further, Figure \ref{fig:domain_nodalDistance} reports the nodal distances $\{d_i\}_{i=1}^{n-1}$ against their corresponding right nodes $\{z_{i+1}\}_{i=1}^{n-1}$, where we recall $d_i \equiv z_{i+1}-z_{i}$. The plot includes several sample eigenfunctions $f_n(r)$ to illustrate the behavior across different excitation indices $n$.

The figure confirms that, for each eigenfunction $f_n(r)$, the nodal distances $\{d_i\}_{i=1}^{n-1}$ increase with radial position, showing modest growth for the initial nodes but substantial increases for the final nodes. 
Moreover, this pattern remains consistent across different eigenfunctions, exhibiting similar behavior as the excitation index $n$ varies.

\begin{figure}[htb]
\centering
     \subfloat[][{Parabolic fit of the outermost node $z_{n}(n)$.}]{
    \includegraphics[scale=0.35]{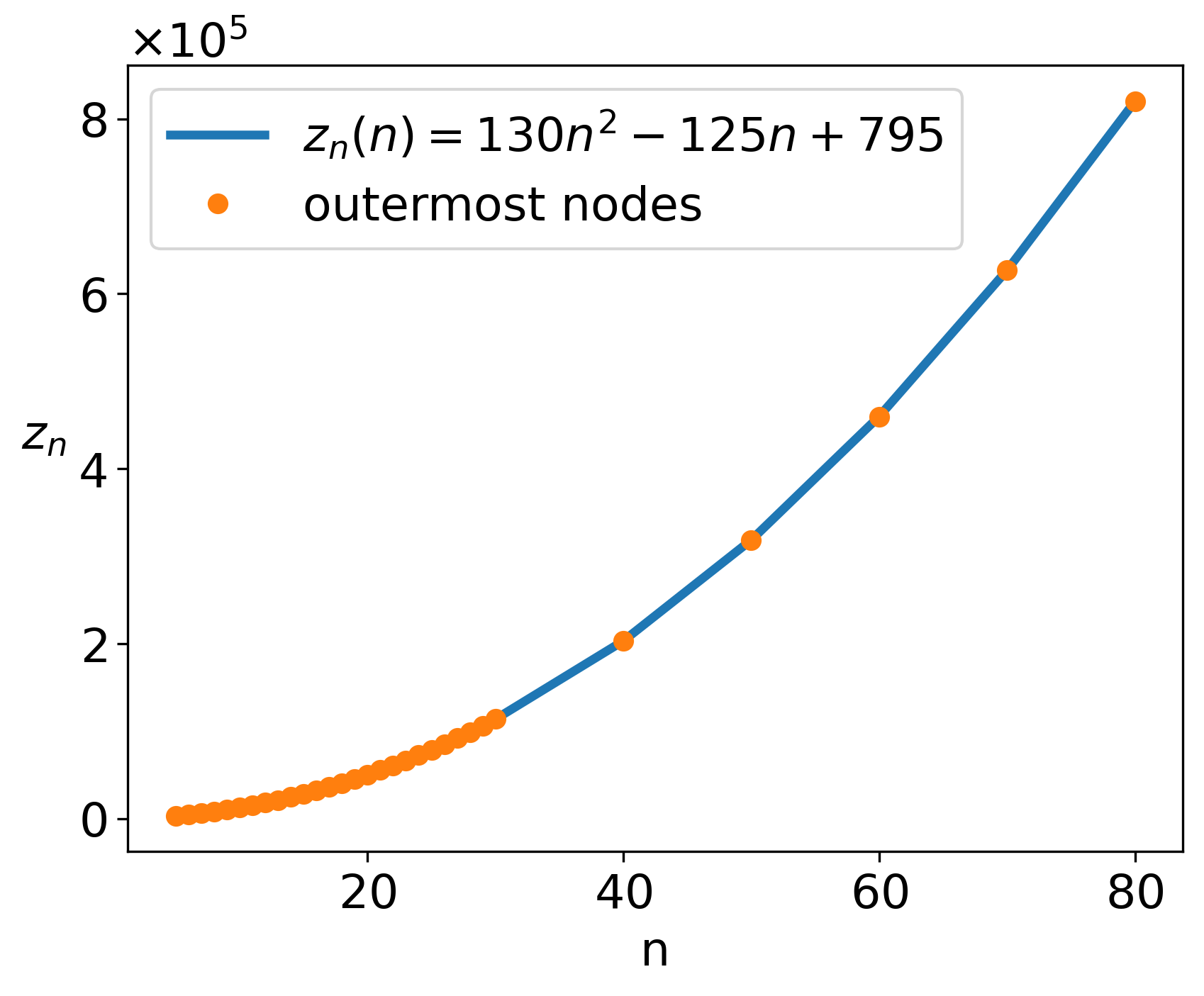}
    \label{fig:domain_rOn}
    } \\
    \subfloat[][{Parabolic fit of the outermost nodal distance $d_{n-1}(n)$.}]{
    \includegraphics[scale=0.35]{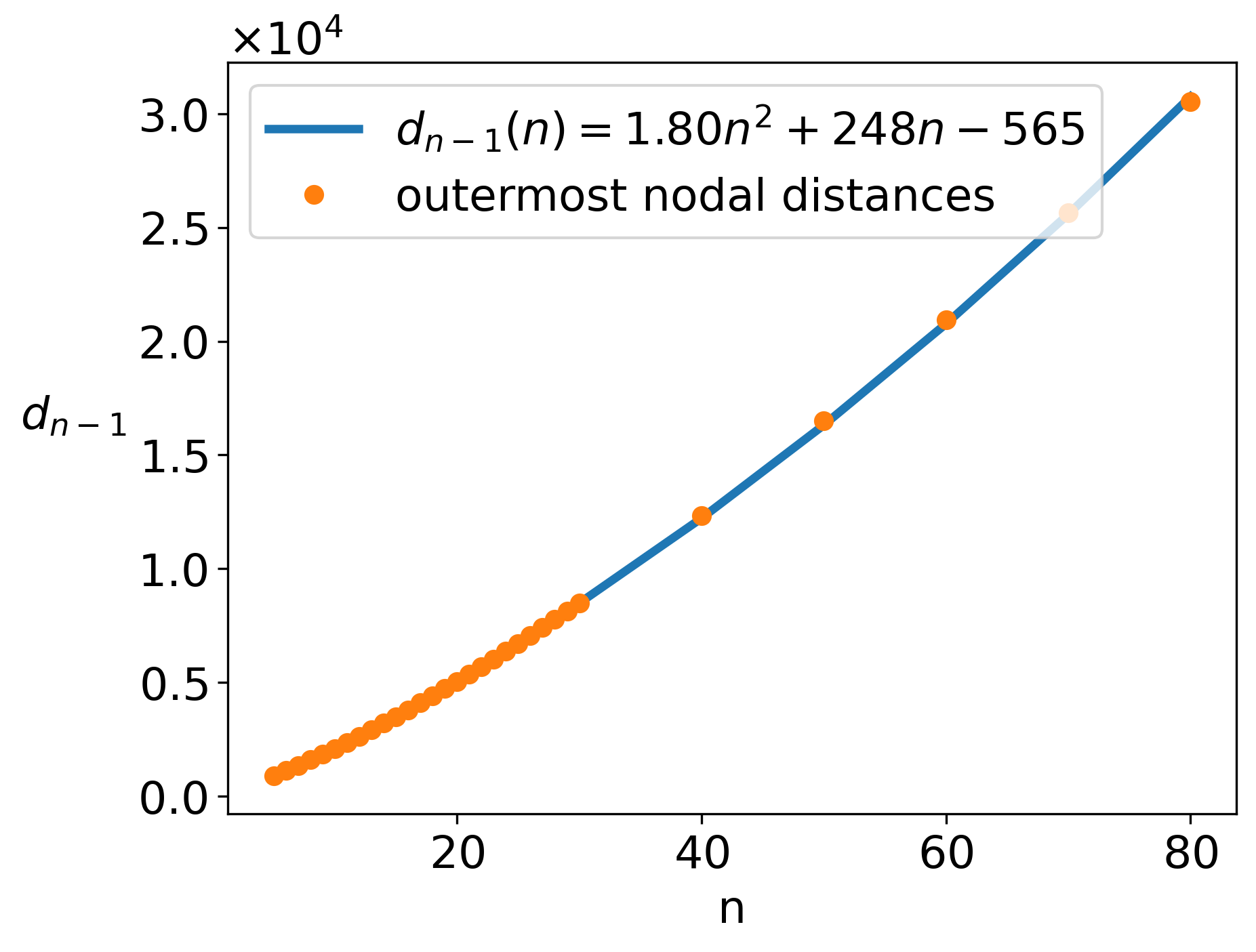}
    \label{fig:domain_ndOn}
    } 
\caption{Variation with excitation index $n$ of (a) the outermost node $z_n(n)$ and (b) the outermost nodal distance $d_{n-1}(n)$ of the eigenfunction. }
\label{fig:domain}
\end{figure}

To examine the regularity of this nodal distance pattern, we seek an appropriate rescaling of the curves in Figure \ref{fig:domain_nodalDistance} that might reveal a common underlying structure across different values of $n$. A natural approach is to normalize the patterns using their outermost point $(z_n, d_{n-1})$, which comprises the outermost node $z_{n}$ and the outermost nodal distance $d_{n-1} \equiv z_n - z_{n-1}$:
\begin{align}
    Z_{i+1} \equiv \frac{z_{i+1}}{z_{n}(n)} \,; && 
    D_i \equiv \frac{d_i}{d_{n-1}(n)} \,;
    && \text{for }i=1,\dots,n-1\,.
    \label{eq:ndRescaling}
\end{align}
For this rescaling to be meaningful and truly capture the fundamental pattern structure, we expect the scaling quantities to exhibit a clear dependence on the excitation index, $z_n(n)$ and $d_{n-1}(n)$.
Testing this hypothesis in Figure \ref{fig:domain}, we find that both quantities follow a distinct parabolic relationship with $n$, which can be quantitatively expressed as:
\begin{subequations}
\label{eq:distNodes}
\begin{align}
    z_n (n) &= 130n^2 -125n + 795 \,;
    \label{eq:nodes} \\
    d_{n-1}(n) &= 1.80n^2 +248n -565 \,.
    \label{eq:dist}    
\end{align}
\end{subequations}

The heuristic laws \eqref{eq:distNodes} validate our choice of the outermost point $(z_n(n), d_{n-1}(n))$ for characterizing the nodal distance pattern. Since this point depends solely on the excitation index, it captures a fundamental property of the eigenfunction and provides an effective basis for the rescaling defined in Equations \eqref{eq:ndRescaling}.

\begin{figure}[htb]
    \centering
    \includegraphics[scale=0.35]{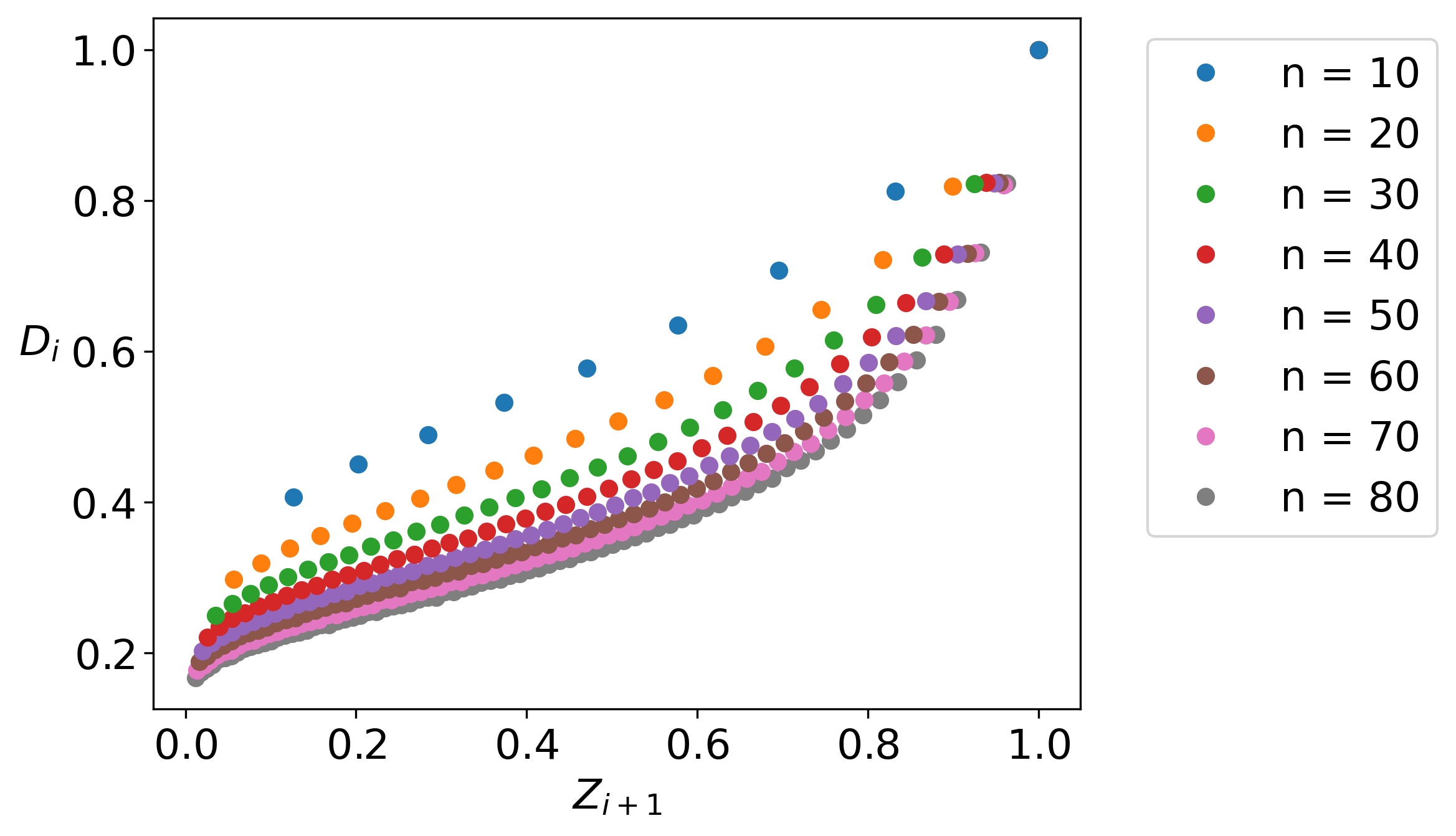}
    \caption{Normalized nodal distances $\{D_i\}_{i=1}^{n-1}$ plotted against their normalized right nodes $\{Z_{i+1}\}_{i=1}^{n-1}$, for several sample eigenfunctions. Normalization follows Equation \eqref{eq:ndRescaling}.}
    \label{fig:domain_nodalDistance_rescaled}
\end{figure}

As shown in Figure \ref{fig:domain_nodalDistance_rescaled}, the rescaled nodal distance patterns, while not perfectly coincident, demonstrate increasing convergence with larger values of $n$, approaching a universal curve in the large $n$ limit. \\

\subsection{ Amplitude modulation}
\begin{figure}[htb]
\centering
    \includegraphics[scale=0.37]{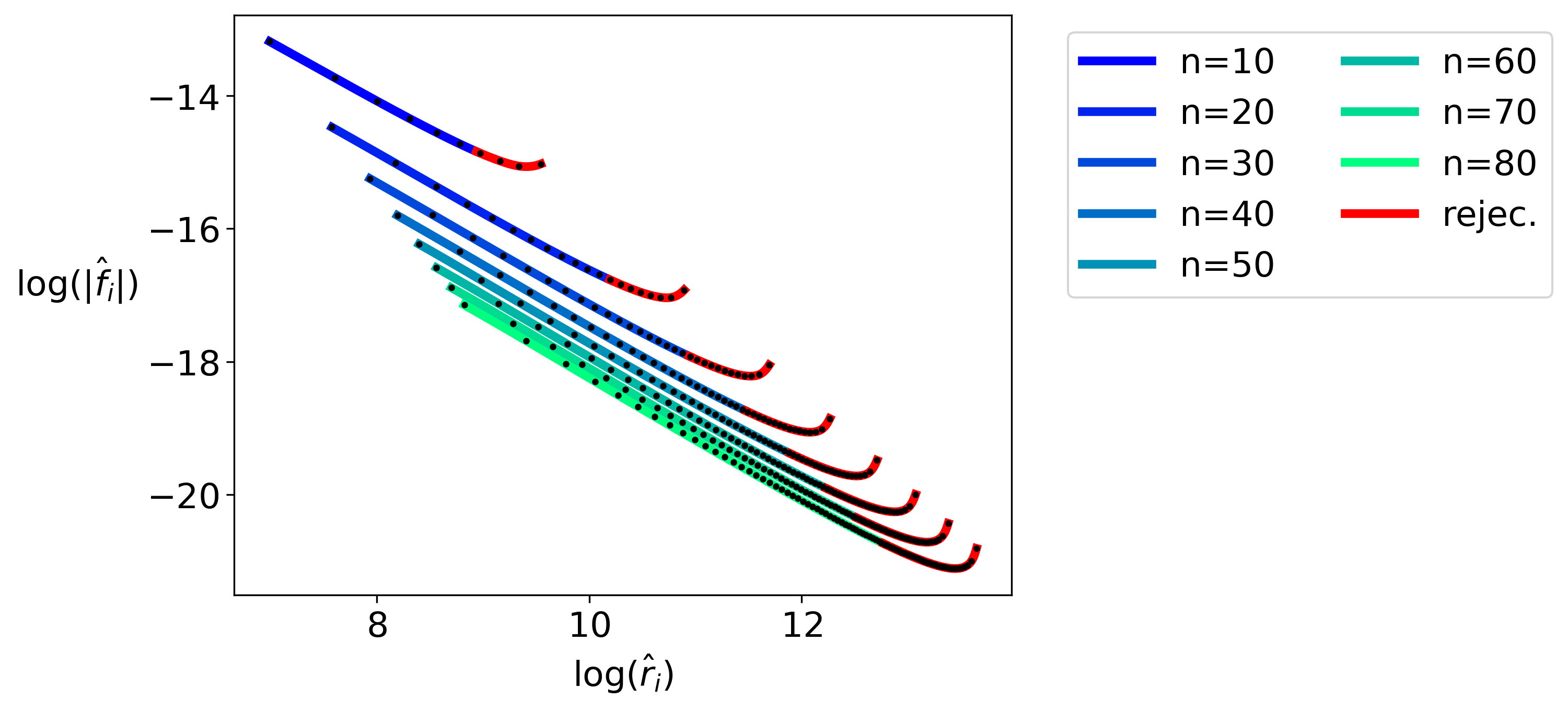}
\caption{Logarithmic plot showing the fitted amplitudes $\{(\hat{r}_i,|\hat{f}_i|)\}_{i=1}^{n}$ of eigenfunction oscillations for sample excitation indices $n$. Red regions indicate points deviating from linear behavior, which are excluded from the fit. }
\label{fig:eigenfunctionFit_loglog}
\end{figure}

Let us now examine the eigenfunction's local extrema $\{(\hat{r}_i,\hat{f}_i)\}_{i=0}^{n}$, whose amplitudes $|\hat{f}_i|$ decrease with increasing $\hat{r}_i$. To formalize this behavior, in Figure \ref{fig:eigenfunctionFit_loglog} we present a logarithmic scale plot of the absolute values of local extrema $\{(\hat{r}_i,|\hat{f}_i|)\}_{i=1}^{n}$ for sample eigenfunctions, excluding the central extremum $(\hat{r}_0,|\hat{f}_0|)$ since its radius $\hat{r}_0=0$ is incompatible with logarithmic scaling. We observe that the inner amplitudes exhibit linear decay, while the outer amplitudes deviate from this behavior, reaching a minimum at a given $r_{min}$ and then increasing again. Based on this, we perform a fit over the region from the first local extremum $\hat{r}_1$ to a fixed distance before the minimum, $0.95r_{min}$. It results that the amplitudes follow a power law, with parameters $a(n)$, $b(n)$ dependent on the excitation index $n$ and with negative exponent $a(n)<0$:
\begin{align}
    |\hat{f}_i| = b(n) \,\hat{r}_i^{\,\,a(n)}\,.
\end{align}

\begin{figure}[htb]
\centering
    \includegraphics[scale=0.37]{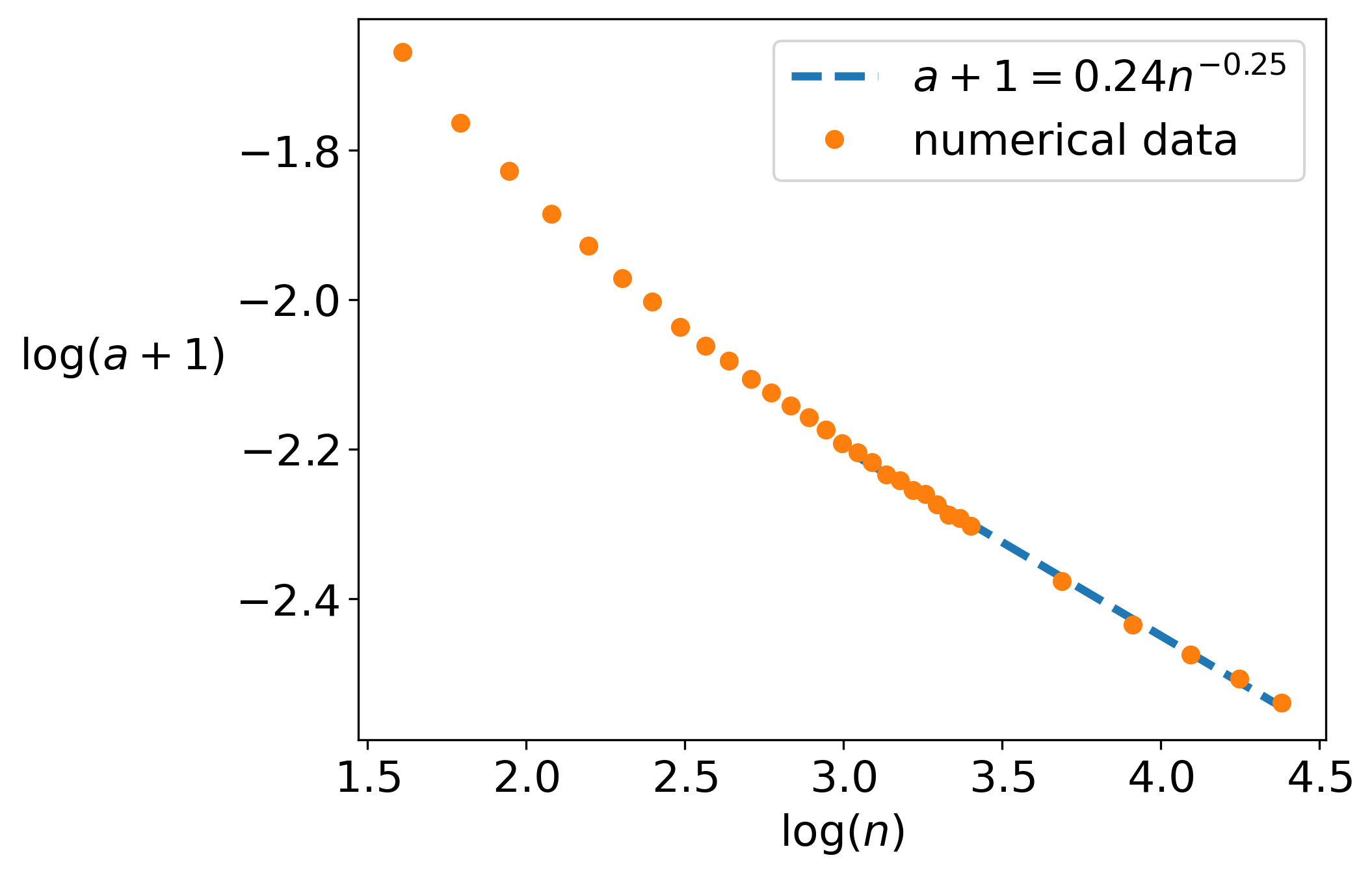}
\caption{Power law fits for the exponential parameter $a(n)=-1+0.24 n^{-0.25}$, in logarithmic scale. }
\label{fig:eigenfunctionFit_alpha}
\end{figure}

Figure \ref{fig:eigenfunctionFit_alpha} shows that the exponent $a(n)$ approximately follows a power law in $n$, approaching the asymptotic value $-1$ in the large $n$ limit. The behavior stabilizes for $n \gtrsim 20$, which is therefore selected as the onset of the fitting region. The resulting heuristic law reads:
\begin{align}
    a(n) &= -1 + 0.24 n^{-0.25} \,.
    \label{eq:varyN_aFit}
\end{align}
Observe that the asymptotic value $a \sim -1$ aligns with the approximate behavior $f_n(r) \sim r^{-1}$ commonly adopted in the literature (\cite{Matos2024,Bohmer2007}) and consistent with the flat plateau of the observed rotation curves.  

\subsection{Eigenvelocities}
\begin{figure}[htb]
\centering
     \subfloat[][{Eigenvelocity profile with extrema.}]{
    \includegraphics[scale=0.35]{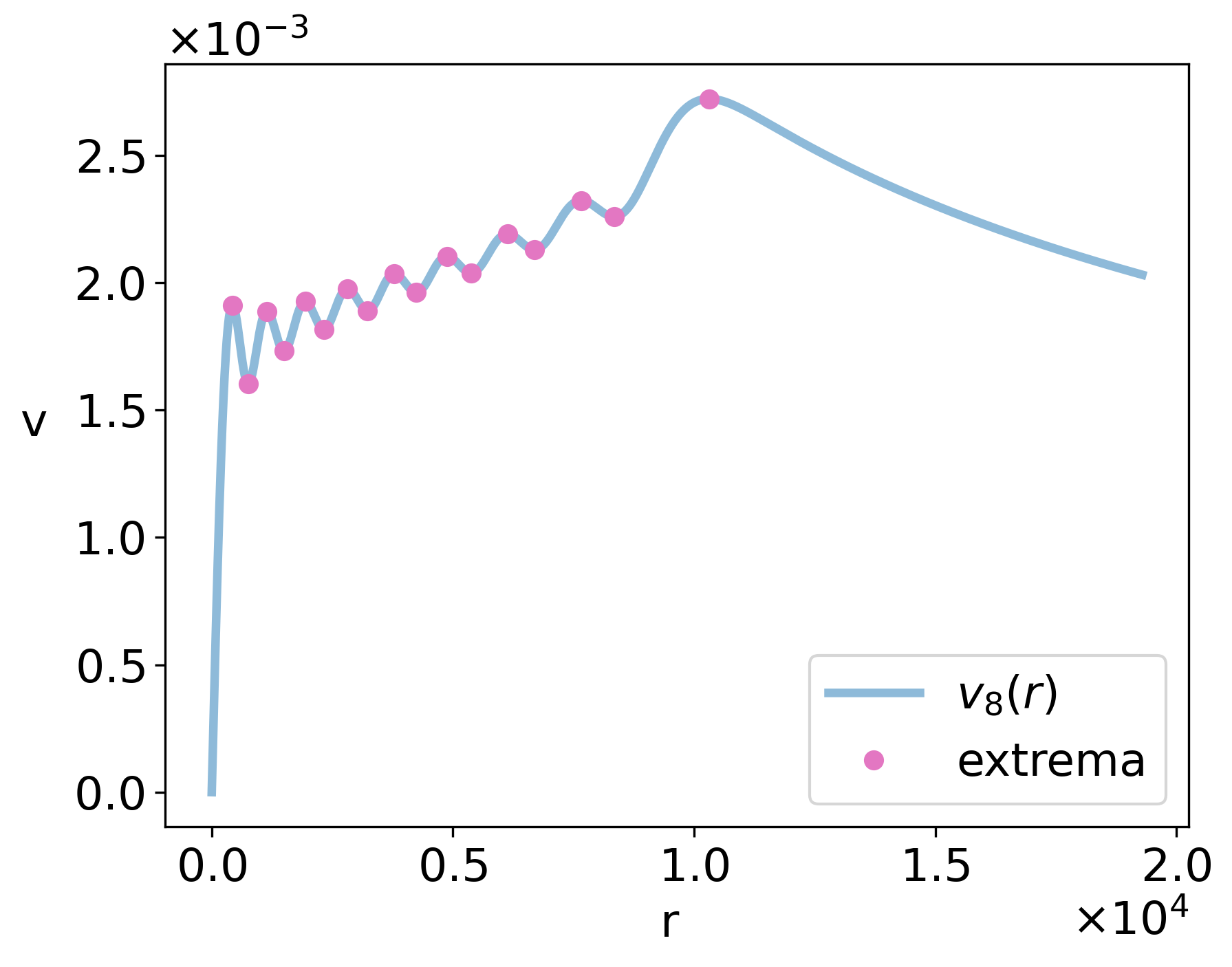}
    \label{fig:vDetails}
    }\\
    \subfloat[][{Magnified view with notation.}]{
    \includegraphics[scale=0.35]{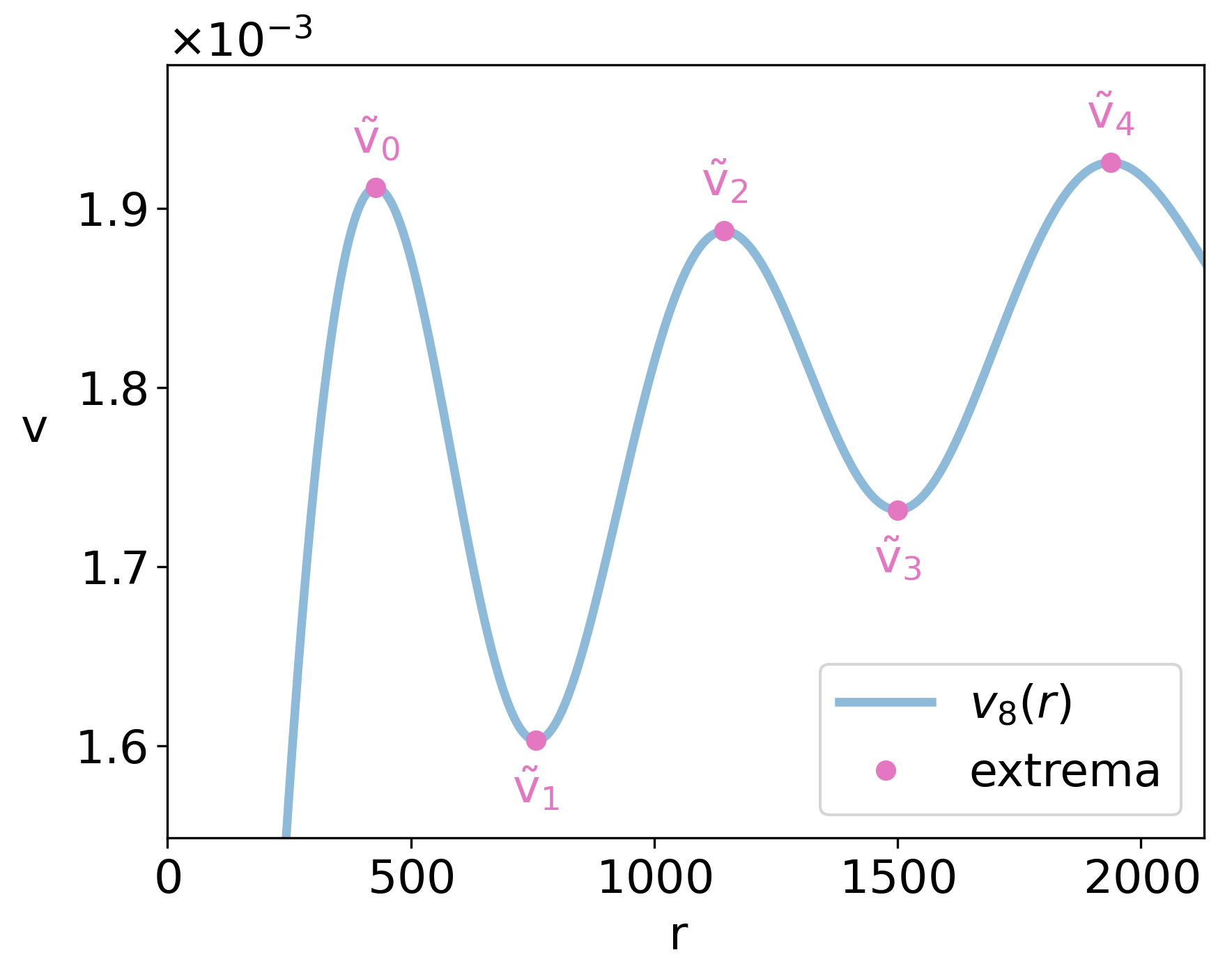}
    \label{fig:vDetails_zoom}
    } 
\caption{Eigenvelocity $v_{8}(r)$ with detailed notation. Panel (a) shows the complete eigenvelocity profile, with local extrema $\{(\tilde{r}_i,\tilde{v}_i)\}_{i=0}^{16}$. Panel (b) provides a magnified view highlighting the notation.}
\label{fig:eigenvelocity_notation}
\end{figure}

Finally, let us examine the eigenvelocities $v_n(r)$  associated with the eigenfunctions $f_n(r)$, according to definition \eqref{eq:velocity}. Figure \ref{fig:eigenvelocity_notation} presents an example, again using a low excitation index $n=8$ for clarity, highlighting the curve's key features.

Similarly to their experimental counterparts, namely the galactic rotation curves, these profiles exhibit three distinct regions: an initial linear raise, a mid-range oscillatory region, and a final Keplerian decline. The oscillatory region contains $2n+1$ local extrema, denoted by $\{\tilde{r}_i, \tilde{v}_i\}_{i=0}^{2n}$ (see Figure \ref{fig:eigenvelocity_notation}).The radial positions $\{\tilde{r}_i\}_{i=0}^{2n}$ of these extrema lie between the eigenfunction's nodes $\{z_i\}_{i=1}^{n}$ and the eigenfunction's local extrema $\{\hat{r}_i\}_{i=0}^{n}$. 

\begin{figure}[htb]
    \centering
    \includegraphics[scale=0.37]{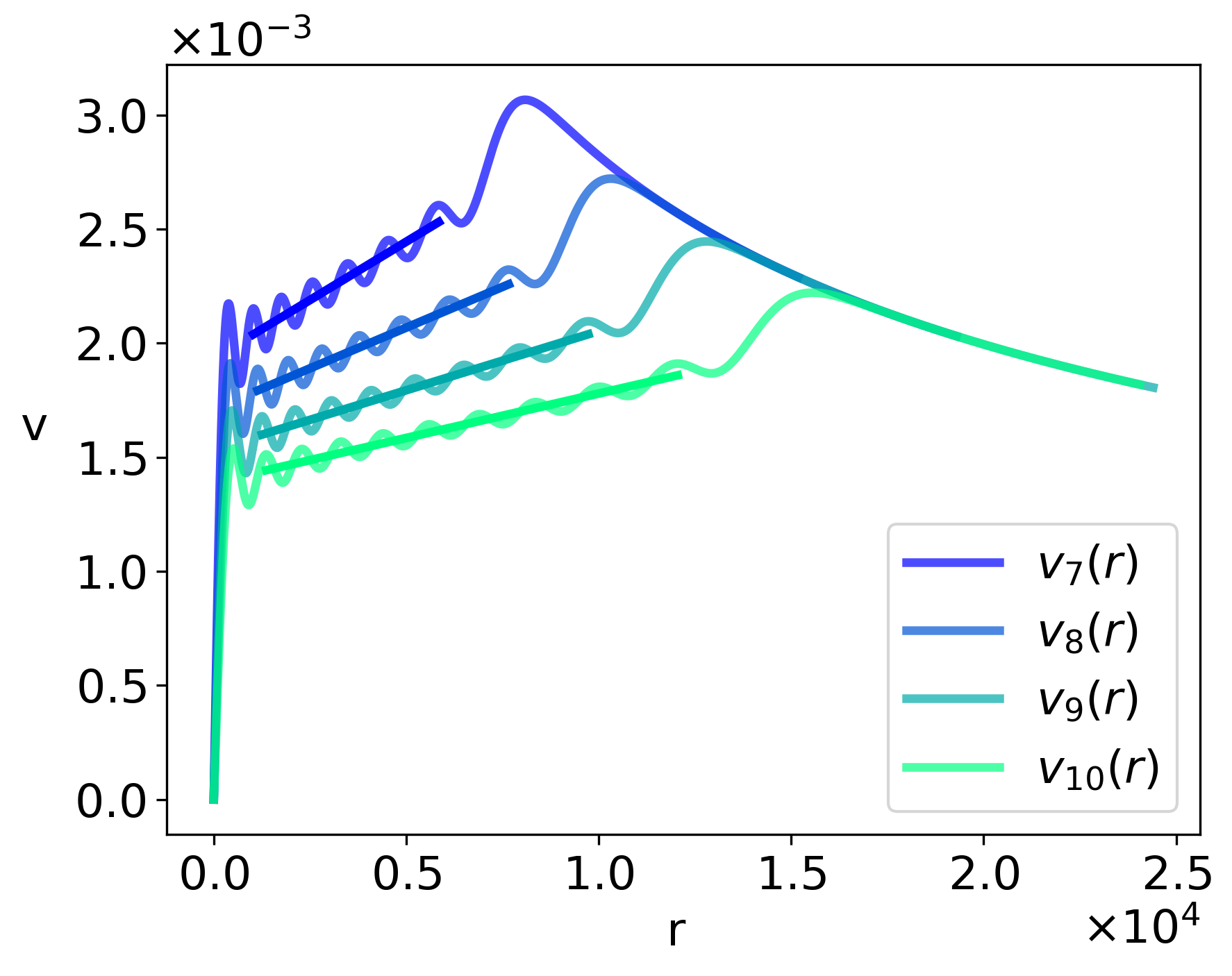}
    \caption{ Examples of eigenvelocity profiles $v_n(r)$ with mid-range linear fit.}
    \label{fig:velocityFit_examples}
\end{figure}

\begin{figure}[htb]
\centering
    \includegraphics[scale=0.37]{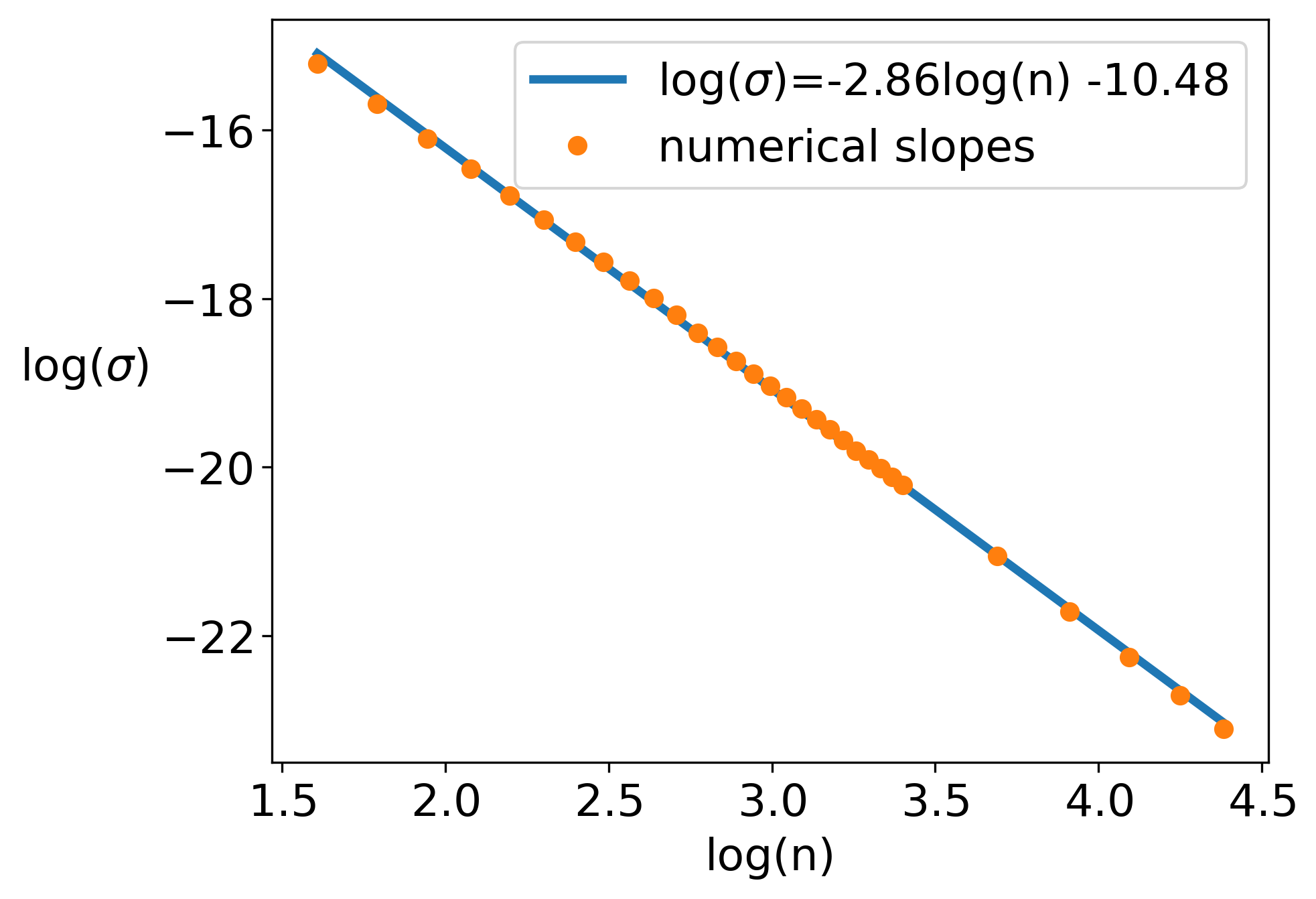}
\caption{Power law relationship between the slopes $\sigma(n)$ and the excitation index $n$, derived from mid-range fits of eigenvelocities (see Eq. \ref{eq:linearOscillating}).}
\label{fig:velocityFit_slopes}
\end{figure}

As shown in Figure \ref{fig:velocityFit_examples} for sample eigenvelocities $v_n(r)$, the oscillating region exhibits a distinctly linear average trend. 
We perform linear fits in the mid-range region, excluding the first two and last two local extrema for improved accuracy:
\begin{align}
    v_n(r) &= \sigma(n) r + q(n) &&
    \text{for } \tilde{r}_2 \le r \le \tilde{r}_{2n-2}  
    \label{eq:linearOscillating}
\end{align}
Figure \ref{fig:velocityFit_slopes} plots the resulting slopes $\sigma(n)$ against the excitation number $n$. These slopes decrease with increasing excitation numbers, according to a power law:
\begin{align}
    \sigma(n) = (2.82\cdot10^{-5}) n^{-2.86}
    \end{align}
In the large $n$ limit the slopes approach zero, consistent with the experimental observations of flat plateaux and with the result we would obtain by plugging in the definition \eqref{eq:velocity} the $f_n(r)\sim r^{-1}$ approximation obtained for the eigenfunctions.  \\
%

\begin{figure}[htbp]
\centering
     \subfloat[][{Power law fit for $\tilde{v}_{2n}(\tilde{r}_{2n})$.}]{
    \includegraphics[scale=0.35]{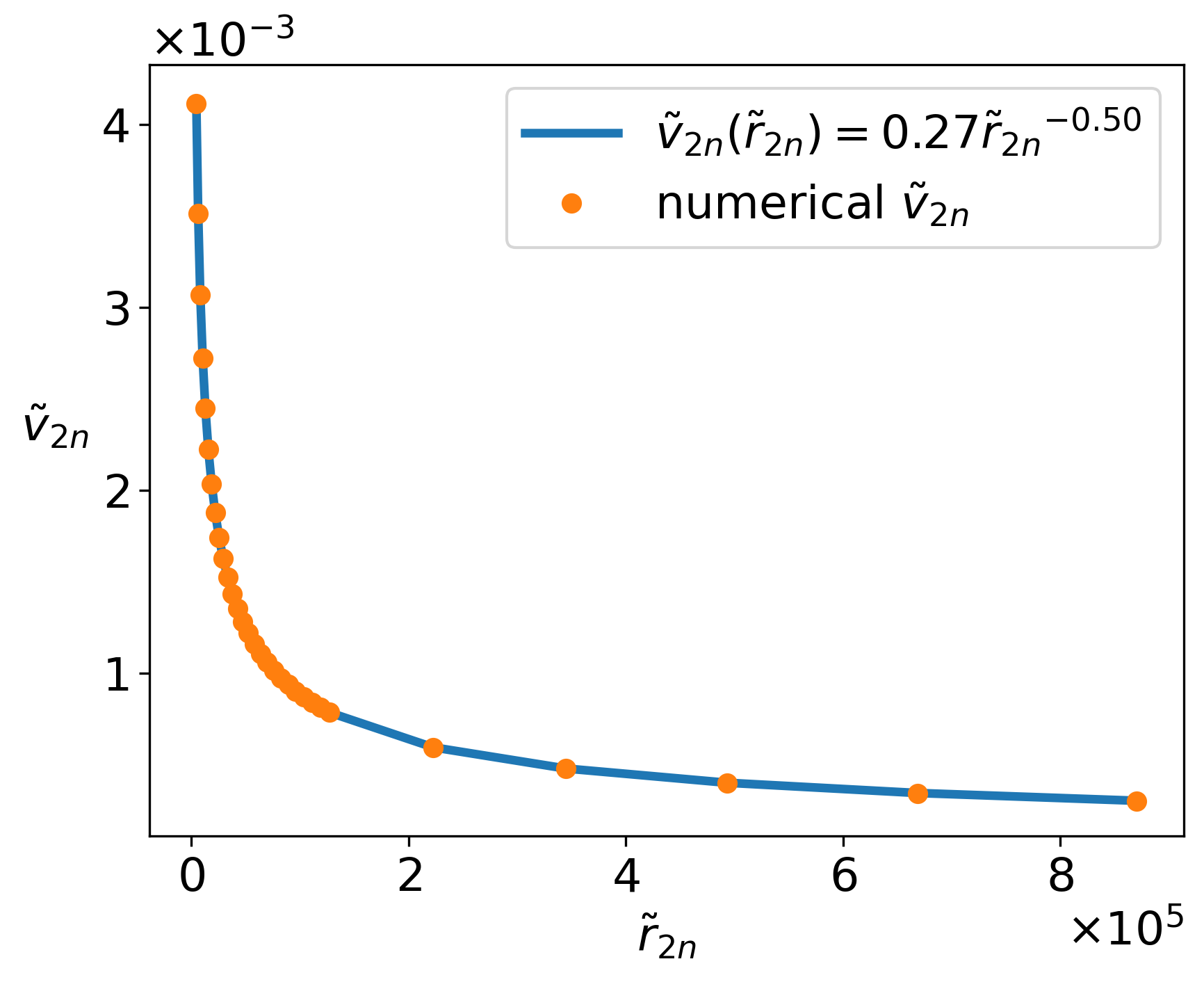}
    \label{fig:velocityFit_vout}
    }\\
    \subfloat[][{Parabolic fit for $\tilde{r}_{2n}(n)$.}]{
    \includegraphics[scale=0.35]{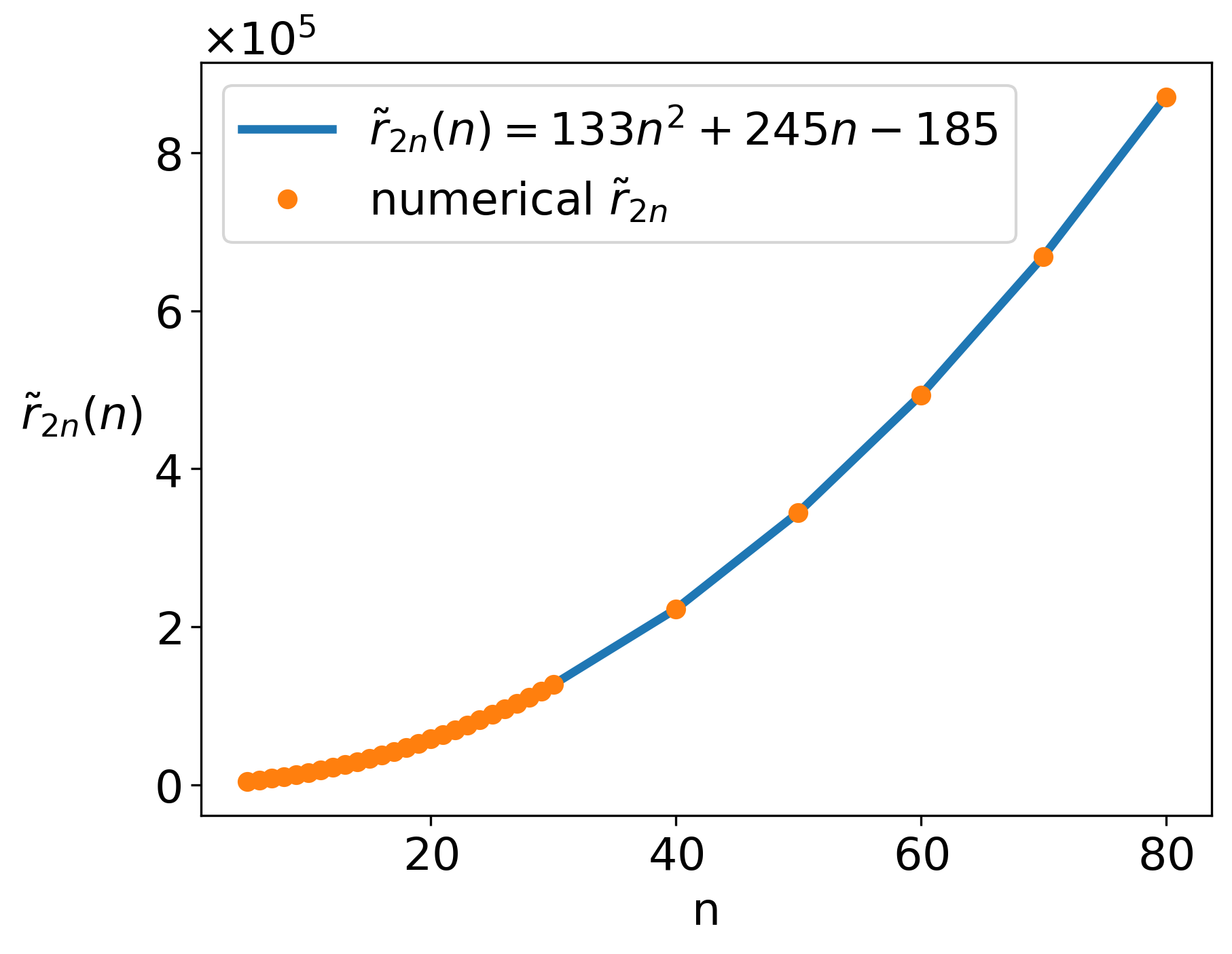}
    \label{fig:velocityFit_rout}
    } 
\caption{Heuristic laws describing the outermost local extremum $(\tilde{r}_{2n}(n),\,\tilde{v}_{2n}(\tilde{r}_{2n}))$ of the eigenvelocity $v_n(r)$. The velocity follows a power law in the radius (see Eq. \eqref{eq:velocityFit_out_v}), while the radius shows parabolic dependence on the excitation index (see Eq. \eqref{eq:velocityFit_out_r}).}
\label{fig:velocityFit}
\end{figure}

To conclude, we analyze the outermost local extremum $(\tilde{r}_{2n},\tilde{v}_{2n})$  of the eigenvelocities. Figure \ref{fig:velocityFit_examples} indicates that its velocity exhibits a clear relationship with the corresponding radius, $\tilde{v}_{2n}(\tilde{r}_{2n})$, while the radius shows a distinct dependence on the excitation index, $\tilde{r}_{2n}(n)$. These relationships, plotted in Figure \ref{fig:velocityFit}, yield the following quantitative heuristic laws:
\begin{subequations}
\label{eq:velocityFit_out}
\begin{align}
    \tilde{v}_{2n}(\tilde{r}_{2n}) &= 0.27 \,\tilde{r}_{2n}^{\,\,-0.5} \,;
    \label{eq:velocityFit_out_v}\\
    \tilde{r}_{2n}(n) &= 133n^2 + 245n -185 \,.
    \label{eq:velocityFit_out_r}
\end{align}
\end{subequations}

The radius $\tilde{r}_{2n}(n)$ exhibits a parabolic dependence on $n$, similar to the behavior observed for $\hat{r}_{n}(n)$ and $z_{n}(n)$ in the eigenfunction. According to formula \eqref{eq:velocityFit_out_v}, the velocity $\tilde{v}_{2n}(\tilde{r}_{2n})$ decays as the square root of the inverse radius, consistent with a Keplerian decline encompassing the entire mass distribution. 

\begin{figure}[htb]
\centering
    \includegraphics[scale=0.32]{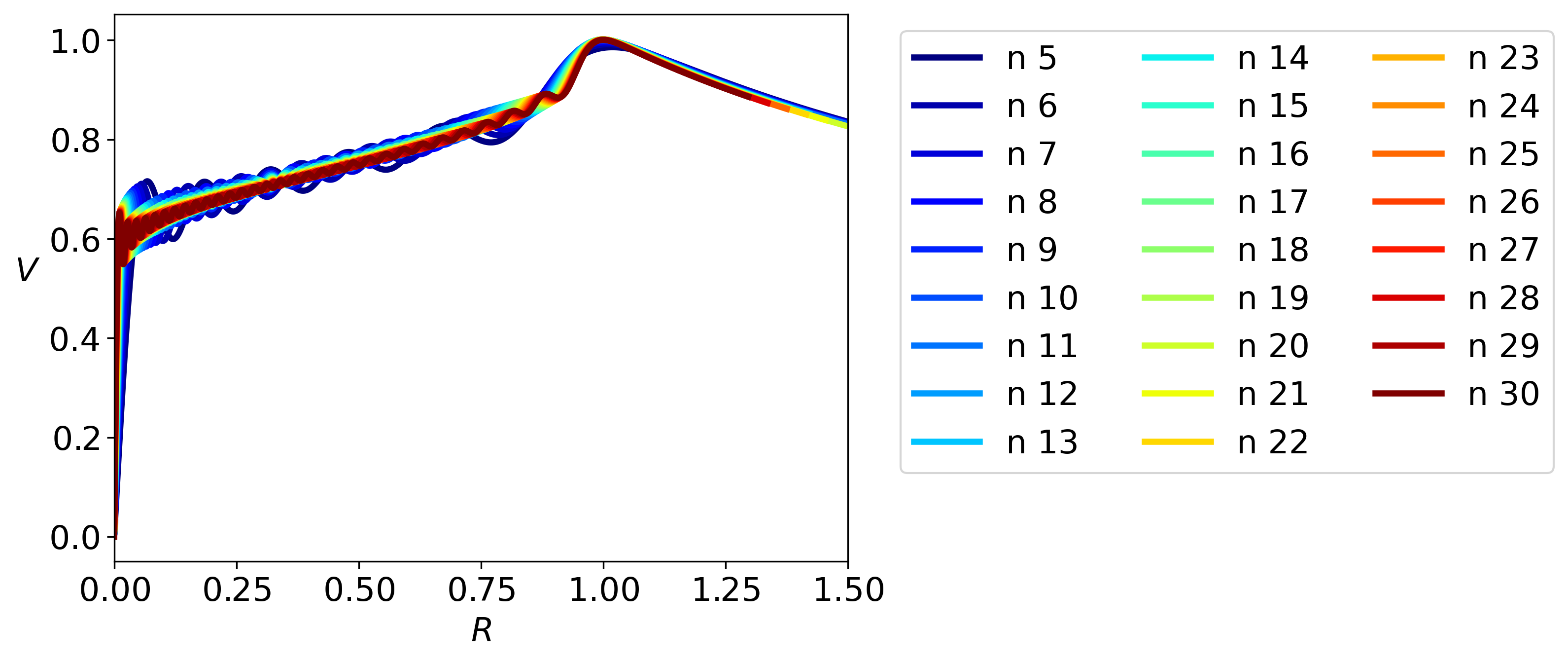}
\caption{Universality in eigenvelocity profiles, after rescaling according to Equations \eqref{eq:velCurve_scaling}.}
\label{fig:universality}
\end{figure}

Overall, the heuristic laws \eqref{eq:velocityFit_out} suggest a natural scaling for the eigenvelocities, that depends solely on the excitation index $n$:
\begin{align}
    R &\equiv \frac{r}{\tilde{r}_{2n}(n)} \,;    &
    V &\equiv \frac{v}{\tilde{v}_{2n}(n)} \,,
    \label{eq:velCurve_scaling}
\end{align}
where $\tilde{v}_{2n}(n) $ is obtained by combining Equations \eqref{eq:velocityFit_out_v} and \eqref{eq:velocityFit_out_r}, $\tilde{v}_{2n}(n) \equiv \tilde{v}_{2n}(\tilde{r}_{2n}(n))$. 
Figure \ref{fig:universality} reports eigenvelocities rescaled according to Equation \eqref{eq:velCurve_scaling}. The plot shows how the numerically computed rotation curves, originally shown in Figure\ref{fig:velocityFit_examples}, collapse onto a single average curve after rescaling, revealing an intrinsic universal behavior. 
%

\section{Discussion and Conclusions} \label{sec:conclusions}
This work provides a characterization of excited Schr\"odinger-Poisson eigenstates, extending the current knowledge of their fundamental structure.
The analysis is based on numerically derived heuristic laws that describe how key properties of the eigenstates scale with the excitation index $n$.

The eigenfunctions are characterized by two main components: oscillatory behavior and amplitude modulation. Unlike standard oscillatory functions, the nodal spacing increases with radial position, showing more pronounced growth at larger radii. The amplitudes decay following a power law, with deviations observed only at the last local extrema, and an exponent approaching $-1$ in the large $n$ limit. This description both aligns with and refines the widely used approximation $f_n(r) \sim \sin(r) r^{-1}$. The heuristic laws also provide insight into the extent of modeled matter distributions, with a support -- approximately defined by the radial position of the last local extremum --  showing a parabolic dependence on  $n$. 

The resulting eigenvelocity characterization confirms and extends the results previously derived in the literature \citep{Sin1994}. The mid-range oscillatory region exhibits a linear trend with a positive slope that decays with $n$ as a power law, approaching zero in the large $n$ limit. The clear dependence on $n$ displayed by the radial position and velocity of the last extremum provides natural scaling laws, through which a universal behavior emerges for the eigenvelocities, intrinsically characterizing the model.

Concerning the instability issue mentioned in the introduction, we
emphasize that the excited eigenstates we study are \textit{unstable}, as proved in \citep{Guzman2006}, and thus they may not be the optimal candidate to directly model galactic dark matter distributions. However, we include a remark on a mathematical peculiarity, framed in a dynamical systems perspective. 
We observe that the instability of a stationary state does not imply unobservability. Indeed, as is well known, a dynamical system can spend a large amount of time in the neighborhood of an unstable (hyperbolic) equilibrium before moving away from it. Regarding dark matter, this possibility has often been considered, starting with the early works of \citet{Sin1994} and \citet{Lee1996}, then with \citet{Guzman2004}, up to more recent works such as \citet{Roque2023}, who observed that 
\quotes{the shortest living unstable mode for each $n$ (i.e., the one with the largest real part) has a lifetime that increases with $n$. In this sense, higher excited states are less unstable than lower excited ones}.
This comment refers to the exponential growth factor $e^{\lambda_n t}$ characterizing instability, with $\lambda_n$ the positive real part of the stability eigenvalues representing the reciprocal of the characteristic instability time $\tau=\lambda_n^{-1}$.
From the numerical simulations reported by \citep{Roque2023} (see Figure 4 therein), the value $\lambda_n$ seems to approach zero for $n\to\infty$, implying that the characteristic instability time $\tau=\lambda_n^{-1}$ may diverge.
The larger the instability time, the more significant the probability that, starting near an excited state, a configuration close to that state can be observed. Applying this argument to support the use of single highly excited eigenstates as models for dark matter is beyond the scope of this work. 
This discussion simply aims to offer a more complete overview of the role of instability, so as to give due weight to the assumptions made on this regard when setting the stability requirements a model should satisfy. In a more general perspective, these considerations just imply that instability is not necessarily a limitation for physical applications. Thus, the comprehensive mathematical characterization of the excited eigenstates presented in this work can still be useful, supporting future analytical and physical developments in different areas. 

Depending on the specific applications, this work could be extended in different directions, for example by including external sources in the Poisson equation \citep{Ji1994} or by adding self-interaction effects \citep{Lee1996}. Examining the dependence on $n$ of the key features in these new frameworks would further clarify how they connect to the mathematical structure of the problem.  
A systematic investigation of this aspect is reserved for future work.


\section*{Acknowledgements}
We thank Massimo Robberto, Lead of NIRCam STScl \& Johns Hopkins University, for valuable discussions on dark matter during his visit to our Department.

We acknowledge GNFM (INdAM), the Italian national group for Mathematical Physics.

We acknowledge ICSC – Centro Nazionale di Ricerca in High Performance Computing, Big Data and Quantum Computing, funded by European Union – NextGenerationEU.

\appendix

\bibliographystyle{elsarticle-num-names} 
\bibliography{bibliography_stationarySP,bibliography_rotationCurves,bibliography_SP}

\end{document}